\documentclass[%
aps,
prl,
superscriptaddress,
floatfix,
citeautoscript,
 amsmath,amssymb,
 reprint,%
]{revtex4-1}
\setcitestyle{super}
\makeatletter
\renewcommand\NAT@biblabelnum[1]{#1\,.}
\makeatother

\usepackage{graphicx}
\usepackage{epstopdf}
\usepackage{dcolumn}
\usepackage{xcolor}
\usepackage{wasysym}
\usepackage{hyperref}
\usepackage[noabbrev, nameinlink]{cleveref}

\graphicspath{{./graphics/}}
\definecolor{tred}{rgb}{0.71,0.16,0.43}

\newcommand{\bcaption}[2]{\caption[#1]{\textbf{#1} #2}}

\begin{document}

\title{Sensing dot with high output swing for scalable baseband readout of spin qubits} 

\author{Eugen Kammerloher}
\affiliation{JARA-Institute for Quantum Information, RWTH Aachen University, 52074~Aachen, Germany}
\author{Andreas Schmidbauer}
\author{Laura Diebel}
\affiliation{Fakultät für Physik, Universität Regensburg, 93040~Regensburg, Germany}
\author{Inga Seidler}
\author{Malte Neul}
\author{Matthias Künne}
\affiliation{JARA-Institute for Quantum Information, RWTH Aachen University, 52074~Aachen, Germany}
\author{Arne Ludwig}
\author{Julian Ritzmann}
\author{Andreas Wieck}
\affiliation{Applied Solid State Physics, Ruhr-Universität Bochum, 44801~Bochum, Germany}
\author{Dominique Bougeard}
\affiliation{Fakultät für Physik, Universität Regensburg, 93040~Regensburg, Germany}
\author{Lars R. Schreiber}
\author{Hendrik Bluhm}
\affiliation{JARA-Institute for Quantum Information, RWTH Aachen University, 52074~Aachen, Germany}
\email[email to: ]{kammerloher@physik.rwth-aachen.de}
\date{\today}
\definecolor{Blue}{rgb}{0,0,1}
\definecolor{Green}{rgb}{0,1,0}
\newcommand{\LS}[1]{\textcolor{Blue}{\small LS: #1}}
\newcommand{\HB}[1]{\textcolor{Green}{\small HB: #1}}

\begin{abstract}
A crucial requirement for quantum computing, in particular for scalable quantum computing and error correction, is a fast and high-fidelity qubit readout.
For semiconductor based qubits, one limiting factor for local low-power signal amplification, is the output swing of the charge sensor.
We demonstrate GaAs and Si/SiGe asymmetric sensing dots (ASDs) specifically designed to provide a significantly improved response compared to conventional charge sensing dots.
Our ASD design features a strongly decoupled drain reservoir from the sensor dot, which mitigates negative feedback effects found in conventional sensors.
This results in a boosted output swing of $3\,\text{mV}$, which exceeds the response in the conventional regime of our device by more than ten times.
The enhanced output signal paves the way for employing very low-power readout amplifiers in close proximity to the qubit.
\end{abstract}

\maketitle

\section*{Introduction}
Spin qubits based on gate-defined quantum dots (QDs) use proximal charge sensors \cite{Field1993, Schoelkopf1998} and, more recently, dispersive gate sensing techniques \cite{Colless2013, West2018, Zheng2019} for readout of the quantum state after spin-to-charge conversion \cite{Elzerman2003, Johnson2005}.
Proximal charge sensors can be quantum point contacts or sensing dots (SDs), with the latter being the most sensitive sensor for spin qubit readout.
The focus of new readout circuits is shifting towards scalability, as high-fidelity, scalable readout is a key requirement for quantum computers with more than just a few qubits.

The state-of-the-art readout technique is based on RF reflectometry \cite{Reilly2007}, which satisfies the requirement of high fidelity and provides the largest bandwidths to date.
Depending on the qubit scaling strategy, RF reflectometry may be the method of choice, as there are solutions to miniaturize at least parts of the necessary RF components \cite{Mahoney2017} or use multiplexing techniques \cite{Hornibrook2014, Hornibrook2015, Schaal2018}, for example, in the context of a crossbar qubit architecture \cite{Li2018}.
However, the overall readout periphery is complex, and necessary RF components are currently on the centimeter scale.

Baseband readout, using optimized transistor circuits in close proximity to the qubit, may prove decisive for scaling strategies where readout and some control functionality are integrated near each qubit, as for example proposed for the spider web array \cite{Boter2022}.
Single-shot readout using a HEMT amplifier has been demonstrated \cite{Vink2007}, and the performance was improved for amplifiers adjacent to the sample \cite{Tracy2016, Curry2019, Mills2022}.
Increasing the SD output signal would allow an even lower total power consumption, thus more simultaneous qubit readouts or a higher readout fidelity using the baseband readout approach, since the power requirement is determined by amplifier gain and sensitivity.
In conventional sensing dots, however, the output swing is limited by negative feedback, due to a large drain capacitance, analogous to the Miller effect in classical electronics \cite{Horowitz2015}.

In this study, we introduce a proximal charge sensor designed specifically to create an asymmetric sensing dot (ASD) by utilizing additional electrodes to sculpt the electrostatic potential of an SD.
Our main goal is to investigate the performance of this novel gate concept for ASD realization.
We perform voltage bias transport measurements, observing a significant reduction in dot and drain capacitive coupling in Coulomb diamonds.
We find qualitatively similar results in remote doped GaAs/(Al,Ga)As and undoped Si/SiGe devices, supporting the universality of the concept.
Additionally, we use the ASD for charge sensing in a nearby qubit-like double quantum dot (DQD) tuned into a multi-electron regime.

By focusing on the realization and investigation of the ASD gate concept, we aim to establish a foundation for further research into the potential of this sensor design for scalable baseband readout quantum circuits. The ASD allows for a trade-off between the expanded footprint of the sensor gate layout, owing to additional gate electrodes, and its boosted signal output. Recent investigations into the qubit shuttling architecture have eased the requirements on the former aspect \cite{PRXQuantum.4.020305}, as demonstrated by Seidler et al. \cite{Seidler2021}.

\section*{ASD Concept}
\begin{figure}[htbp]
    \includegraphics[width=\linewidth]{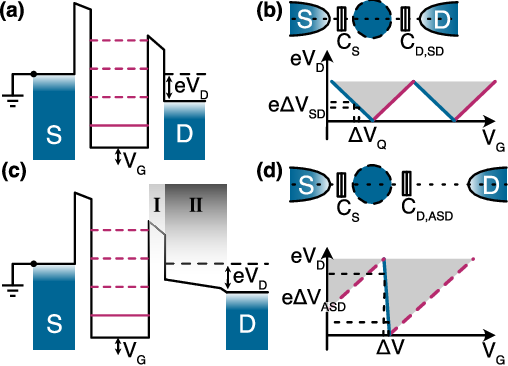}
    \bcaption{Electric potential of SD and ASD and output signal.}
    {\textbf{(a),(c)} show a schematic electric potential with an applied bias of $V_{D}$ between source S and drain D for the SD and ASD, respectively.
The ladder of the dots' levels can be energetically shifted by the gate voltage $V_G$.
    In the SD (panel (a)), the drain D is separated from the dot by a sharp tunneling barrier.
    In the ASD (panel (c)), the barrier is a compound barrier with a sharp tunnel barrier region~\textbf{I} controlling tunneling and an additional intermediate barrier region~\textbf{II} termed slide.
    \textbf{(b),(d)} Schematic Coulomb diamonds for both sensors.
No current flow occurs in the white regions due to Coulomb blockade.
Single-electron transport current flows in the light gray regions.
Blue and red lines mark the edges of the diamond-shaped Coulomb blockade regions.
An equivalent shift $\Delta V_Q$ on the $V_G$-axis results in an enhanced shift $e\Delta V_\text{ASD}$ for the ASD compared to $e\Delta V_\text{SD}$, in current bias mode.
Top insets schematically show a top-view of the potential.
Source and drain electron reservoirs have a tunnel and capacitive coupling to the quantum dot (blue circle).}
    \label{fig:concept}
\end{figure}

\Cref{fig:concept}(a) illustrates the electric potential of a biased SD.
The ladder of QD energy levels can be shifted by the gate voltage $V_G$, while a bias $V_D$ is applied to the drain\,D.
Measuring the transport current through the device while driving these voltages produces a characteristic diamond-shaped pattern due to Coulomb blockade effects.
A section of these diamonds is depicted in panel \cref{fig:concept}(b), where current flows only in the gray regions.
According to a constant interaction model \cite{Hanson2007}, the positive and negative slopes of the Coulomb diamonds are $C_G/(C_{\Sigma}-C_D)$ and $-C_G/C_D$, where $C_{\Sigma}$ is the dot's total capacitance to ground, and $C_G$ and $C_D$ are the capacitance to the swept gate electrode and drain reservoir, respectively.
In the SD, the drain\,D is separated from the dot by a sharp tunneling barrier, similar to the source\,S, as depicted by the top sketch in \cref{fig:concept}(b).

The ASD concept presented in this study is focused on modifying the drain barrier such that the tunneling rate remains comparable to the conventional SD, while significantly reducing the drain capacitance to increase the maximum output voltage.
To achieve this, the barrier is subdivided into a compound barrier with a sharp tunnel barrier region~\textbf{I} and an additional slowly decreasing barrier region~\textbf{II} termed slide.
Region~\textbf{II} physically separates the drain reservoir further from the dot, resulting in $C_{D,ASD}\ll C_{D,SD}$, while region~\textbf{I} controls the tunneling at the working point.
The top sketch \cref{fig:concept}(d) shows the new SD configuration, where the asymmetric arrangement of source and drain is visible, hence the name ASD.

\Cref{fig:concept}(c) displays the electric potential of a biased ASD.
An equivalent shift $\Delta V_Q$ on the $V_G-$axis now results in a greatly enhanced shift $e\Delta V_\text{ASD}$ for the ASD compared to $e\Delta V_\text{SD}$.
Therefore, an essential figure of merit for the ASD is the magnitude of the blue slope $-C_G/C_D$ in \cref{fig:concept}(d), where larger values indicate a higher output voltage swing of the sensor when configured for charge sensing.

\section*{Experimental Results}
\begin{figure}
    \includegraphics[width=\linewidth]{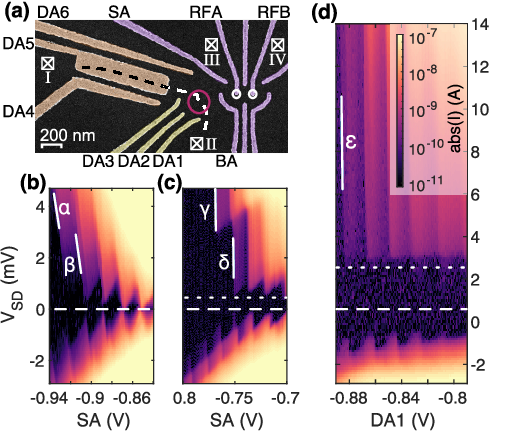}
    \bcaption{Device layout and Coulomb diamond comparison in GaAs/(Al,Ga)As.}
    {\textbf{(a)} False-colored scanning electron micrograph of a GaAs ASD device.
Labels \XBox\,I and \XBox\,II mark the drain and source reservoirs, connected by a dashed line, representing a possible electron trajectory through the sensor.
A red circle indicates the sensor dot position.
White circles denote the position of a DQD, which may host qubits.
    \textbf{(b)} Conventional SD with nearly symmetric source and drain reservoirs is tuned, with gates DA(4-6) turned off (0\,V).
The white dashed horizontal line signifies zero bias.
White lines indicate negative slopes $\alpha=-0.25\,\text{V/V}$ and $\beta=-0.30\,\text{V/V}$ of a linear fit to the $100\,\text{pA}$ contour.
    \textbf{(c)} The ASD is tuned up with an asymmetric source drain configuration using gates DA(4-6).
The white dashed horizontal line denotes zero bias, and the dotted line shows a bias window $0\leq V_{SD}\lesssim 360\,\mu\text{V}$ with no current flow due to the high drain barrier opacity.
White lines represent negative slopes $\gamma=-3.96\,\text{V/V}$ and $\delta=-2.29\,\text{V/V}$ of a linear fit to the $100\,\text{pA}$ contour.
    \textbf{(d)} ASD Coulomb diamonds with a nearby tuned-up DQD for charge sensing.
Gate DA1 is used instead of SA to avoid mis-tuning the DQD, which is less sensitive to this gate electrode.
The white dashed horizontal line marks zero bias, and the dotted line indicates a bias window $0\leq V_{SD}\lesssim 2.9\,\text{mV}$ with no current flow.}
    \label{fig:device_GaAs}
\end{figure}

\Cref{fig:device_GaAs}(a) displays a false-colored scanning electron microscope (SEM) image of a double quantum dot (DQD) integrated with an ASD on the left.
The ASD is incorporated into our GaAs qubit design, utilizing a doped MBE-grown GaAs/(Al,Ga)As heterostructure, which features a 2DEG 90\,nm beneath the interface (see supplemental material for heterostructure details) \cite{Macleod2015,Cerfontaine2020}.

The yellow-colored gates DA(1-3) define an SD, while the orange-colored gates DA(4-6) control the dot-drain transition.
The purple-colored gates are used to define and manipulate the nearby DQD.
The source reservoir (labeled \XBox\,II) has a common distance from the SD (red circle) used for conventional SDs, while the drain reservoir (\XBox\,I) is asymmetrically formed.
A dashed line illustrates a possible electron path connecting the sensor's source and drain.
Supported by numerical electrostatic simulations, we developed device designs optimizing the drain barrier potential region for a slow and monotonically declining transition to the drain reservoir, effectively creating a micron-scale electron slide with sharp tunnel barriers defining the SD \cite{Kammerloher:841219}.
Transport at the optimal working point is still possible, while the capacitive coupling of dot and drain is significantly reduced.

To experimentally validate the ASD concept, we compare Coulomb diamonds of the sensor tuned for conventional operation with nearly symmetric source and drain barriers (gates DA(4-6) set to $0\,\text{V}$) in \cref{fig:device_GaAs}(b), to a tuned up ASD in \cref{fig:device_GaAs}(c).
The sensor is operated in the multi-electron Coulomb blockade regime with a voltage bias $V_{D}$ applied to ohmic contact \XBox\,II (drain), while current is measured with a transimpedance amplifier at \XBox\,I (source).

Typical values of $V_{D}$ do not exceed a few hundred $\mu$V for conventional SD operation.
In the following, values of $V_{D}$ can easily exceed several mV, since the compound drain barrier of the ASD becomes only transparent at higher bias voltages.
\Cref{fig:device_GaAs}(b) shows typical Coulomb diamonds, when operating the device in a conventional way without using the additional gate electrodes, that are symmetric around zero bias.
Gate SA not only shifts the dot minimum, but also modifies the tunneling rates of the dot barriers, thus the diamond features become smeared out for higher tunneling rates at more positive voltages.
At more negative voltages transport is blocked close to zero bias, as the barriers become opaque.
The slope of the diamond can be evaluated at different current levels.
Here we choose $100\,\text{pA}$, which is a compromise between low current operation and a sufficient signal-to-noise ratio.
In \cref{fig:device_GaAs}(b) we fit a line at the $100\,\text{pA}$ contour and find a maximum slope of $\beta=-0.30\,\text{V/V}$ for the SD.
Note, that for a conventional SD the slope remains nearly constant even at higher bias values.

\Cref{fig:device_GaAs}(c) shows equivalent measurements with the sensor tuned to the ASD regime.
The Coulomb diamonds change distinctively from the conventional case and the positive and negative bias configurations are not symmetric anymore.
A bias dependence of the Coulomb lines is observed for slopes $\gamma$ and $\delta$ of the ASD, compared to $\alpha$ and $\beta$ for the SD.
At higher bias, a maximal slope of $\gamma=-3.96\,\text{V/V}$ at the $100\,\text{pA}$ contour is observed\footnote{On a side note: Since $dV_{SD}/dV_{SA}>1$, the ASD has voltage gain and can be used as an ultra low-power, low stray capacitance voltage amplifier in principle, with the limitation of a narrow dynamics range.}.
The steepening of Coulomb diamonds indicates the desired reduction of $C_D$ by a factor of $C_{D,SD}/C_{D,ASD}\approx 13$, compared to the maximal slope in \cref{fig:device_GaAs}(b).
Additionally, we observe a bias window in \cref{fig:device_GaAs}(c), where transport is blocked ($0 \leq V_{SD} \leq V_T$, where $V_T\lesssim 360\,\mu\text{V}$), in between the dashed and dotted white lines.
We find in \cref{fig:device_GaAs}(d) that when a nearby DQD is also tuned up, $V_T$ can be of the order of several mV, which results in low visibility of the typical Coulomb blockade features near zero bias.
Further details of ASD tuning and characteristic features are discussed in the supplemental material.

To test the adaptability of our ASD concept, we also realized a device in a MBE-grown Si/SiGe heterostructure, featuring a $10$\,nm thick Si QW separated from the interface by a $35$\,nm thick $\text{Si}_{65}\text{Ge}_{35}$ spacer \cite{Wild2012} (see supplemental material for heterostructure details). Transferring the ASD from a remote-doped GaAs to an undoped Si/SiGe heterostructure requires a full redesign of the gate pattern:
First, Si/SiGe requires an accumulation gate also above the slide region and second the gate pattern has to be shrunken down to compensate for the three times larger effective electron mass.
Therefore, we implemented a new simulation-guided design to host the ASD and a nearby DQD.
\begin{figure}
    \includegraphics[width=\linewidth]{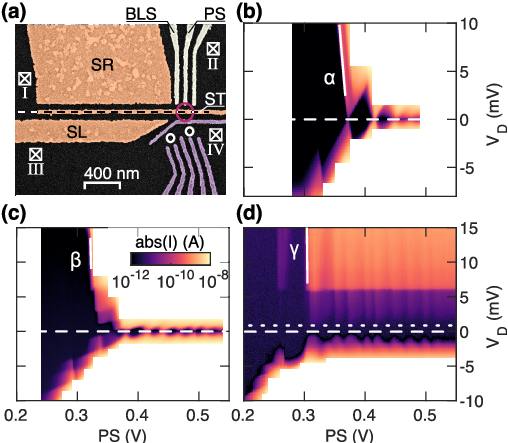}
    \bcaption{Device layout and Coulomb diamond comparison in Si/SiGe.}
             {\textbf{(a)} False-colored SEM image of an ASD device in Si/SiGe similar to the one measured.
              A global top-gate isolated by $\text{Al}_2\text{O}_3$ is not shown.
              The light yellow-colored gates (including BLS and PS) define the sensor QD in vicinity to the purple-colored gates, which can be used to form a DQD.
              The potential slide for the ASD operation of the SD is formed by the orange colored gates (SR, SL and ST).
              The drain and source reservoirs are labeled \XBox\,I and \XBox\,II for the sensor current path and \XBox\,III and \XBox\,IV for the DQD current path.
              \textbf{(b)}-\textbf{(d)} Coulomb diamond measurement series tuning the SD from a symmetric to an asymmetric configuration.
              To do so the gate potential of SR is step-wise decreased from 0.34\,V in (b) to 0.24\,V in (c) to 0.215\,V in (d) and simultaneously the gate potential of BLS is adjusted to maintain the same tunnel rate for transport trough the SD.
              The measurements are performed by varying the drain potential $V_{\text{D}}$ at the reservoir\,\XBox\,I while keeping the reservoir \XBox\,II at zero volt for different potentials at PS, which serves as plunger gate for the sensor quantum dot levels.
              The white lines indicate the negative slopes of the corresponding Coulomb diamonds. Linearly fitting the $50\,\text{pA}$ contour leads to slopes of $\alpha=-0.68\,\text{V/V}$ for (a), $\beta=-3.2\,\text{V/V}$ for (b) and $\gamma=-8.0\,\text{V/V}$ for (d).
             }
    \label{fig:device_SiGe}
\end{figure}
A false-colored SEM image of this gate layout is shown in \cref{fig:device_SiGe}(a).
The sensor dot is formed by the light yellow colored gates, from which gate PS serves as the plunger gate for the sensor quantum dot levels.
Gates SL, SR and ST (colored orange) form the potential slide (region~\textbf{II} in \cref{fig:concept}(c)) and the purple colored gates may be used to tune two tunnel-coupled QDs in the vicinity of the sensor dot.
Current through the sensor is defined by ohmic contacts \XBox\,I and \XBox\,II, while contacts \XBox\,III and \XBox\,IV serve as electron reservoirs for the DQD.
Tuning the electrostatic potential to form a sensor QD below gate PS, we record a series of Coulomb blockade measurements to test the tunability of the slide potential region~\textbf{II}.
Starting from a conventional symmetric configuration in \cref{fig:device_SiGe}(b), the slide is then activated by only reducing the voltage on gate SR, while simultaneously increasing the voltage on gate BLS to retain the same tunnel rate (keep region~\textbf{I} in \cref{fig:concept}(b) constant) for transport through the sensor.
In \cref{fig:device_SiGe}(b)-(d), we decreased the voltage on gate SR from 0.34\,V to 0.24\,V and 0.215\,V, respectively.
Similar to the observation in GaAs, for the most negative SR configuration a blockade region with a threshold voltage of $V_\text{T} = 1\,\text{mV}$ emerges\footnote{Even lower voltages on SR led to yet higher $V_\text{T}$.}.

The series in \cref{fig:device_SiGe}(b)-(d) clearly shows that the negative Coulomb diamond edges become steeper, the smaller the voltage applied to the gate SR.
For all three diamonds shown here, we determined these negative Coulomb diamond slopes, by fitting the $50\,\text{pA}$ contour, which represents the best compromise between a low-current operation of the device and a sufficient signal-to-noise ratio.
The slopes in \cref{fig:device_SiGe}(b)-(d) represent the steepest slopes which we found for each configuration\footnote{Interpolation of the data sets along the plunger gate (PS) direction was used to compensate for the lower measurement resolution on this gate.}.
Starting from $\alpha=-0.68\,\text{V/V}$ for the most symmetric configuration, we reach $\beta=-3.2\,\text{V/V}$ for the intermediate configuration and finally $\gamma=-8.0\,\text{V/V}$ for the most asymmetric configuration, corresponding to a twelve-fold reduction of $C_{D,ASD}$.

Hence, we have shown that the ASD concept is equally efficient across material platforms.
Integrating ASDs into a DQD environment, we have demonstrated a controllable reduction of $C_{D,ASD}$ both in Si/SiGe and GaAs while being able to maintain a tunneling rate useful for operation of the sensor quantum dot.
In the proof-of-principle experiments discussed in \cref{fig:device_GaAs,fig:device_SiGe}, we reached a maximal reduction of $C_{D,ASD}$ by a factor of 13 in GaAs and a factor of 12 in Si/SiGe, implying an increase by the same factor of the voltage swings produced by these ASDs, compared to conventional SD operation.

\section*{ASD charge sensing}
We demonstrate charge sensing operation with the ASD by defining a DQD in the center of the GaAs device, using the purple colored gates shown in \cref{fig:device_GaAs}.
For charge sensing, it is crucial to reconfigure the ASD for current bias to take advantage of the high output voltage swing.
In this case, a constant current source supplies 500\,pA through the drain, while the source is grounded.
We tune the ASD to a sensitive position and record a charge stability diagram of the DQD using gates RFA and RFB, as shown in \cref{fig:sensing}(a) (see supplemental material for tuning details).

\begin{figure}
\includegraphics[width=\linewidth]{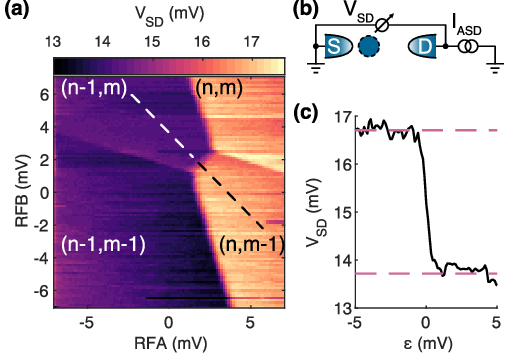}
\bcaption{Charge sensing with current biased ASD.}
    {\textbf{(a)} The ASD is configured for charge sensing of a nearby multi-electron DQD with a $I_{ASD}=500,\text{pA}$ current bias.
The charge stability diagram is recorded with the ASD, using the voltage drop $V_{D}$ across the sensor as the signal (median adjustment per scan-line).
    \textbf{(b)} Schematic of the ASD wiring.
    \textbf{(c)} Cut on the detuning axis $\epsilon$, which corresponds to the dashed line on the left panel (linear background subtracted).
A voltage swing of $3,\text{mV}$ is observed.
}
\label{fig:sensing}
\end{figure}

The voltage drop $V_{D}$ across the ASD is monitored with a voltmeter, as shown in \cref{fig:sensing}(b).
\Cref{fig:sensing}(c) displays the voltage swing across the inter-dot transition of the DQD (dashed line in \cref{fig:sensing}(a)), corresponding to a qubit state change when the DQD is used as an ST-qubit.
We observe a voltage swing of 3\,mV.
The voltage swing is expected to be an order of magnitude larger than that of a conventional SD in this configuration due to the one order of magnitude difference in slopes $\beta$ and $\gamma$ seen in \cref{fig:device_GaAs}. These slopes provide a measure of sensor sensitivity, as discussed in \cref{fig:concept}(c)-(d).

\section*{Discussion}
In conclusion, we have presented a novel type of proximal charge sensor specifically designed for the formation of ASDs as the main goal of this study.
Our findings showcase the successful implementation of an ASD gate concept that is adaptable across various material systems.
This is evidenced by the drain capacitance reduction by factors of 12 and 13 respectively in both undoped Si/SiGe and doped GaAs devices when compared to conventional SD operation.
It is worth noting that the GaAs device serves as a demonstrator, and we anticipate that the ASD is more likely to be relevant for Si-based spin qubit architectures, where nuclear fluctuations play a less significant role and CMOS processing techniques can be employed.
The next experimental step involves integrating the ASD with published work on transistors \cite{Vink2007, Tracy2016, Curry2019} and examining the extent to which the ASD affects the host qubit, considering the higher energy electrons generated during sensor operation compared to conventional SDs \cite{Granger2012, Harbusch2010a}.

\section*{Data availability}
The data sets generated and/or analyzed during this study are available from the corresponding author upon reasonable request.

\section*{Acknowledgements}
This work was funded by ARO under the contract NO W911NF-17-1-0349 titled "A scalable and high performance approach to readout of silicon qubits" and by the German Research Foundation (DFG) within the project BO 3140/4-1.
The device fabrication has been done at HNF - Helmholtz Nano Facility, Research Center Juelich GmbH \cite{Albrecht2017}.

\section*{Competing interests}
The authors declare no competing interests.

\section*{Author contributions}
The study was conceived by H.B.\, L.R.S.\ and D.B.
The GaAs heterostructures were prepared by J.R.\, A.L.\ and A.W.\ and sample fabrication performed by M.K.
E.K.\ conducted the experiments data analysis and numerical simulations on the GaAs samples, advised by H.B.
The Si/SiGe heterostructures were prepared by D.B.
Numerical simulations, the device design and the sample fabrication for Si/SiGe was performed by I.S.\, M.N.\ and L.R.S.
A.S.\, L.D.\ and D.B.\ conducted the experiments and the data analysis on the Si/SiGe samples.
All authors discussed the results.
E.K.\, A.S.\, L.D.\, D.B.\, L.R.S.\ and H.B.\ wrote the manuscript.

\bibliography{bibliography} 
\end{document}


\title{Supplementary Material: Sensing dot with high output swing for scalable baseband readout of spin qubits} 

\author{Eugen Kammerloher}
\affiliation{JARA-Institute for Quantum Information, RWTH Aachen University, 52074~Aachen, Germany}
\author{Andreas Schmidbauer}
\author{Laura Diebel}
\affiliation{Fakultät für Physik, Universität Regensburg, 93040~Regensburg, Germany}
\author{Inga Seidler}
\author{Malte Neul}
\author{Matthias Künne}
\affiliation{JARA-Institute for Quantum Information, RWTH Aachen University, 52074~Aachen, Germany}
\author{Arne Ludwig}
\author{Julian Ritzmann}
\author{Andreas Wieck}
\affiliation{Applied Solid State Physics, Ruhr-Universität Bochum, 44801~Bochum, Germany}
\author{Dominique Bougeard}
\affiliation{Fakultät für Physik, Universität Regensburg, 93040~Regensburg, Germany}
\author{Lars R. Schreiber}
\author{Hendrik Bluhm}
\affiliation{JARA-Institute for Quantum Information, RWTH Aachen University, 52074~Aachen, Germany}
\email[email to: ]{kammerloher@physik.rwth-aachen.de}
\date{\today}

\maketitle

\section*{GaAs/(Al,Ga)As and Si/SiGe heterostructure}
The GaAs/(Al,Ga)As device was fabricated on a MBE-grown heterostructure (sample B14722).
The layer stack is grown at a constant temperature of 695\,$^\circ$C.
A $35\,\text{nm}$ Al$_{0.65}$Ga$_{0.35}$As layer is grown on the GaAs substrate, followed by an additional $50\,\text{nm}$ Si modulation doped Al$_{0.65}$Ga$_{0.35}$As layer.
A $5\,\text{nm}$ Si doped GaAs cap finalizes the structure, thus the 2DEG is formed $90\,\text{nm}$ below the interface.
The ohmic contacts to the 2DEG are thermally activated by a rapid anneal at 460\,$^\circ$C of a gold/germanium alloy.
The mobility of the 2DEG, as obtained by Hall measurements at a temperature of $4.2\,\text{K}$, is of the order of $1.47 \times 10^6\,\text{cm}^2/\text{(Vs)}$ at an electron density of $1.82 \times 10^{11}\,\text{cm}^{-2}$.
A single layer of metal gates is fabricated by means of electron beam lithography.

The Si/SiGe device was fabricated on a solid-source MBE-grown heterostructure.
A relaxed virtual substrate consists of a graded buffer grown at 500\,$^\circ$C up to a composition of Si$_{0.65}$Ge$_{0.35}$ on a Si substrate without intentional miscut and a layer of constant composition Si$_{0.65}$Ge$_{0.35}$.
It provides the basis for a 10\,nm natural Si QW grown at a substrate temperature of 350\,$^\circ$C.
The QW is separated from the interface by 35\,nm Si$_{0.65}$Ge$_{0.35}$.
The structure is protected by a 1.0\,nm naturally oxidized Si cap.
The implanted ohmic contacts to the QW are thermally activated by a rapid anneal at 700\,$^\circ$C.
The mobility of the 2DEG formed in the QW, as obtained by Hall measurements at a temperature of $1.5\,\text{K}$, is of the order of $1.1\times 10^{6}\,\text{cm}^2/\text{(Vs)}$ at an electron density of $6.6 \times 10^{11}\,\text{cm}^{-2}$ and is limited by remote impurity scattering.
A 10\,nm layer of Al$_2$O$_3$ grown by atomic layer deposition insulates the first metal gate layer and the underlying heterostructure.
The metal gates are fabricated by means of electron beam lithography.
A second gate layer, insulated from the first gates by 50\,nm of Al$_2$O$_3$, is used to induce a two-dimensional electron gas in the QW via the field effect.
A Co magnet is deposited on top of the device.
Bias-cooling using voltages of $-0.5$\,V at the gates of the first metal layer and $-4.8$\,V applied to global gate of the second metal layer, is employed for the Si/SiGe devices from room temperature to 300\,mK, to inhibit leakage between the gate layers.

\section*{Characteristic features of the ASD Coulomb diamonds}
\begin{figure}
    \includegraphics[width=\linewidth]{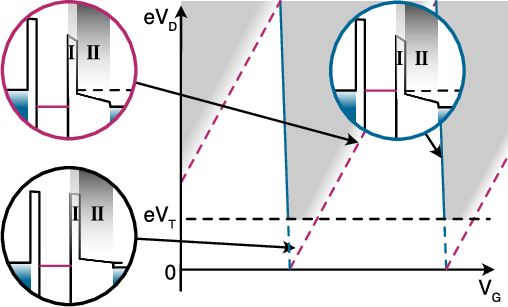}
    \bcaption{Characteristic features of ASD Coulomb diamond.}
             {Schematic of Coulomb diamonds expected for the ASD.
              Single electron current flows in the light gray regions separated by a blue solid line and red dashed line from Coulomb blockade regions (white).
              The white current-free region is extended by the threshold voltage $V_T$ (black dashed line).
              Three insets in black, red and blue frames, depict details of the ASD potential at the corresponding ($V_G$, $V_D$) operation points marked by arrows.
             }
    \label{fig:details}
\end{figure}

In \cref{fig:details}, we discuss several features of the ASD Coulomb diamonds in more detail.
We focus on $V_D>0$, since the ASD has a preferential bias direction and the behavior for $V_D<0$ is unimportant for the intended mode of operation.
Regions in the Coulomb diamond plot where transport is possible, are colored in gray.
The blue slope is the desired working point for readout and the corresponding chemical potentials of the reservoirs and last occupied energy level of the QD are depicted in the blue inset.
The red energy level in the QD is close to the chemical potential of the source.
Tunneling through the thin barrier~\textbf{I} is possible at sufficiently high bias, after which the electron passes through region~\textbf{II} and relaxes down into the drain reservoir.
Along the opposing diamond edge, the energy level in the dot is close to the chemical potential of the drain, as illustrated in the red inset.
However, tunneling through barrier~\textbf{I} in addition to barrier~\textbf{II} becomes exponentially difficult, hence current flow degrades when approaching the dashed red slope from the left.
The black inset depicts low bias configurations, below the dashed black line.
The inset shows a state where the dot energy level is inside the bias window and can potentially contribute to the current flow, however tunneling is nearly impossible, due to the combined barrier thickness of regions~\textbf{I}-\textbf{II}.
The compound drain barrier is only transparent above a threshold $V_T$, when a sufficiently large gradient is formed in region~\textbf{II} by the bias voltage.
Since the ASD compound drain barrier is exceedingly sensitive to gate voltage changes on nearby structures, compensation for these changes can reduce the operational bias voltage space to even higher values of $V_T$.
Additionally, any disorder effects that create local minima in region~\textbf{II}, may result in an even higher value of $V_T$ to compensate.

\section*{ASD Tuning}
The ASD can be tuned under voltage or current bias.
Tuning under voltage bias involves slope evaluation of Coulomb diamond features, to find the desired regime.
Tuning under current bias is closer to the intended mode of operation and lends itself for routine use and automation, since the output voltage swing can be directly measured and easily optimized by a parameter sweep.

\begin{figure}
    \includegraphics[width=\linewidth]{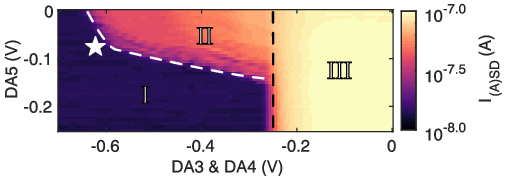}%
    \bcaption{Slide gate vs.\ barriers diagram.}
             {This measurement reveals several distinct features in a voltage bias ($V_{D}=100\,\mu\text{V}$) measurement by sweeping gates DA(3-5).
              Region~III, at voltages where the barriers do not suppress the 2DEG below the gates is undesired leakage.
              Region~II, where a channel exists below DA5.
              Some line structure can be observed, indicating a disorder dot situated in the channel.
              In region~I, current flow is fully suppressed.
              A white star marks a useful voltage starting point for tuning.
             }%
    \label{fig:sld2wall}
\end{figure}

In the following, the tuning of the GaAs device is discussed under voltage bias initially.
For an untuned device, gate electrodes SA and BA should be set below their pinch-off value and provide a fixed barrier, so current can only flow from \XBox\,I to \XBox\,II.
Gate DA6 is set to the same voltage as gate SA, and can be used later, for fine-tuning the sensor and compensate for unrelated tuning operations performed on the nearby qubit.
A useful starting point for ASD tuning, is the slide gate electrode (DA5) versus slide barrier gate electrodes (DA3 and DA4, while DA6 is kept fixed) diagram in \cref{fig:sld2wall}.
Two parallel conducting channels are observed.
Conduction below DA5 (marked with II) and undesired leakage below DA3 or DA4 far away from the ASD (III).
The dashed lines represent the conduction onset for these channels.
Gate voltages marked with a white star provide starting values for DA(3-5), when tuning the ASD.
Here, the lever arms of DA(3-5) are similar, indicating a barrier formation close to the desired first sensor barrier between DA3 and DA6.
Some line structure can be observed, indicating a disorder dot situated in the channel.
Simulations suggest, that a dot can arise between DA5 and DA3, at certain gate voltage combinations, even without an additional disorder potential, due to the gate geometry.
The second sensor barrier is defined in the following step.

\begin{figure}
    \includegraphics[width=\linewidth]{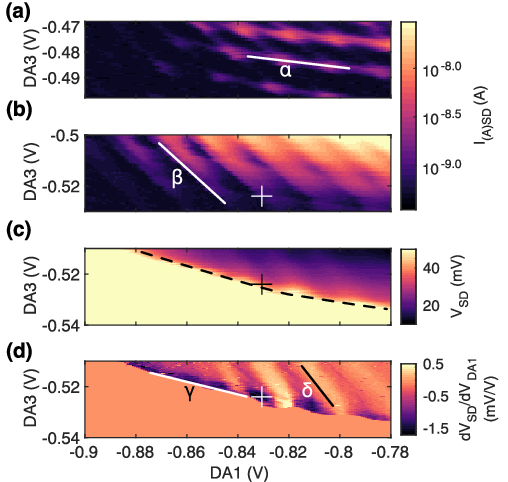}
    \bcaption{Sensor wall to wall diagrams.}
    {\textbf{(a)} At a voltage bias of $V_{D}=100\,\mu\text{V}$, the sensor dot has a disorder dot in series, which further modulates the current Coulomb blockade.
     Slope $\alpha$ is $-0.13\,\text{V/V}$.
     \textbf{(b)} At a voltage bias of $V_{D}=1\,\text{mV}$, the disorder dot is nearly suppressed, revealing the sensor's Coulomb lines.
     Slope $\beta$ is $-0.95\,\text{V/V}$.
     \textbf{(c),(d)} Sensor is reconfigured for current bias of $I_{ASD}=500\,\text{pA}$ and $V_{D}\leq 50\,\text{mV}$ compliance, with a otherwise unchanged gate voltage configuration.
     \textbf{(d)} The derivative in sweep-direction reveals the Coulomb line structure and areas with large voltage swing.
     The maximum voltage swing is marked with a cross.
     Slope $\gamma$ and $\delta$ are $-0.26\,\text{V/V}$ and $-1.10\,\text{V/V}$.
    }
    \label{fig:tuning}
\end{figure}

Sweeping DA1 versus DA3 reveals Coulomb lines of the sensing dot (\cref{fig:tuning}(a),(b)).
At a low voltage bias of $V_{D}=100\,\mu\text{V}$ in \cref{fig:tuning}(a), the sensor has a disorder dot in series, which further modulates the Coulomb blockade.
Raising the bias voltage to $1\,\text{mV}$ in \cref{fig:tuning}(b), nearly suppresses the disorder dot.
The strong influence of $V_{D}$ indicates the disorder dot's position to be in the channel below DA5 or between DA5 and DA3.
To further tune the sensor under voltage bias mode, we measure Coulomb diamonds.
The negative diamond slope can be evaluated and optimized for a higher magnitude at different settings of DA(1-6).
However, finding suitable gate voltages is a very device layout specific process, due to the strong capacitive interdependence of the sensor gates and may require several parameter sweeps, while evaluating line features in Coulomb diamond plots at each iteration.
In the following, the sensor is tuned further in current bias mode, which is the more expedient approach.

The sensor is now reconfigured for current bias of $I_{ASD}=500\,\text{pA}$ with an voltage compliance of $V_{D}\leq 50\,\text{mV}$ and the same parameters otherwise.
In \cref{fig:tuning}(c) the voltage drop across the sensor is depicted.
The voltage drop is of the order of tens of mV for this sensor configuration and saturates at the compliance value below the dashed line, when transport is fully blocked.
The derivative in sweep-direction in \cref{fig:tuning}(d) reveals a structure, similar to the Coulomb lines in \cref{fig:tuning}(b), where slope $\beta$ is similar to $\delta$.
Gate DA3 has a strong influence on the transport, since it controls the compound drain barrier.
We observe regions of maximum voltage swing close to the full blockade (one such region marked with white cross), which is the desired working point of the ASD for charge sensing.
Voltage changes on nearby structures can shift the working point and can require compensation on the ASD gates, since the sensitive regions are small in extent.

\section*{Measurement setup}
\begin{figure}[ht]
    \includegraphics[width=\linewidth]{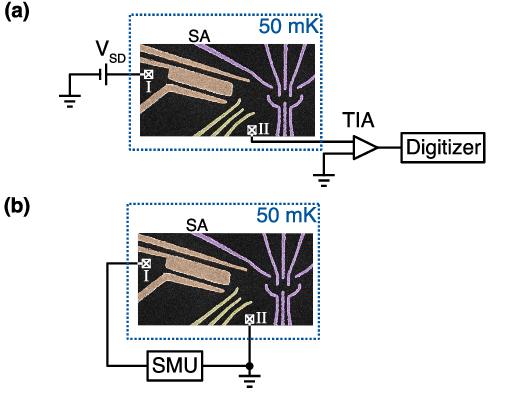}
    \bcaption{Measurement setup.}
             {\textbf{(a)} Schematic diagram of the voltage bias setup and the GaAs SEM image.
             \textbf{(b)} Schematic diagram of the current bias setup. The source measure unit (SMU) is configured for current bias and measures the voltage $V_{SD}$ across the device.}
    \label{fig:setup}
\end{figure}
Here we briefly present the measurement setups used in this work.
We show the GaAs case only, with only minor variations for the SiGe setup.
\Cref{fig:setup}(a),(b) depict the schematic for the voltage and current bias cases respectively.
In \cref{fig:setup}(a) a Basel Instruments SP983c TIA is used to measure current through the SD.
All voltage signals including the SD bias $V_{SD}$ are generated using our homebuild DecaDAC voltage source.
In \cref{fig:setup}(b) a Keithley 2400 source measure unit is used to supply a constant current and measure the resulting voltage swing $V_{SD}$.